\documentclass[fleqn,usenatbib]{mnras}

\usepackage{newtxtext,newtxmath}

\usepackage[T1]{fontenc}
\usepackage{ae,aecompl}

\usepackage{graphicx}	
\usepackage{amsmath}	
\usepackage{amssymb}	
\usepackage{longtable}
\usepackage[caption=false]{subfig}
\usepackage{multirow}

\title[Weak high-redshift radio quasars]{Unveiling the weak radio quasar population at $z\geq4$}
\author[K. Perger et al.]{Krisztina Perger,$^{1,2}$\thanks{E-mail: k.perger@astro.elte.hu}
 S\'andor Frey,$^{2}$ Krisztina \'E. Gab\'anyi,$^{3,2}$, L. Viktor T\'oth$^1$
\\
$^{1}$Department of Astronomy, E\"{o}tv\"{o}s Lor\'{a}nd University, P\'{a}zm\'{a}ny P\'{e}ter s\'{e}t\'{a}ny 1/A, H-1117 Budapest, Hungary\\
$^{2}$Konkoly Observatory, Research Centre for Astronomy and Earth Sciences, Konkoly Thege Mikl\'{o}s \'{u}t 15-17, H-1121 Budapest, Hungary\\
$^{3}$MTA-ELTE Extragalactic Astrophysics Research Group, P\'{a}zm\'{a}ny P\'{e}ter s\'{e}t\'{a}ny 1/A, H-1117 Budapest, Hungary}

\date{Accepted 23 September 2019. Received 30 August 2019; in original form 7 June 2019}

\pubyear{2019}

\begin{document}
\label{firstpage}
\pagerange{\pageref{firstpage}--\pageref{lastpage}}
\maketitle%


\begin{abstract}
We applied image stacking on empty-field Faint Images of the Radio Sky at Twenty-Centimeters (FIRST) survey maps centred on optically identified high-redshift quasars at $z\geq4$ to uncover the hidden $\mu$Jy radio emission in these active galactic nuclei (AGN). The median stacking procedure for the full sample of $2229$ optically identified AGN uncovered an unresolved point source with an integrated flux density of 52~$\mu$Jy, with a signal-to-noise ratio $\sim10$. We co-added the individual image centre pixels to estimate the characteristic monochromatic radio power at $1.4$~GHz considering various values for the radio spectral index, revealing a radio population with $P_\mathrm{1.4GHz}\sim10^{24}$~W~Hz$^{-1}$. Assuming that the entire radio emission originates from star-forming (SF) activity in the nuclear region of the host galaxy, we obtained an upper limit on the characteristic star formation rate, $\sim4200$~M$_\odot$~yr$^{-1}$. The angular resolution of FIRST images is insufficient to distinguish between the SF and AGN origin of radio emission at these redshifts. However, a comparison with properties of individual sources from the literature indicates that a mixed nature is likely. Future very long baseline interferometry radio observations and ultra-deep Square Kilometre Array surveys are expected to be sensitive enough to detect and resolve the central $1-10$~kpc region in the host galaxies, and thus discriminate between SF and AGN related emission.

\end{abstract}



\section{Introduction}
Since the discovery of the first  quasar at redshift $z\geq4$ \citep{1987Natur.325..131W}, the number of active galactic nuclei (AGN) known at the highest redshifts is continuously increasing. As a result of extensive observing campaigns and surveys \citep[e.g.,][]{2000AJ....120.1579Y,2010AJ....140.1868W,2011AJ....142...72E, 2016arXiv161205560C,2016MNRAS.460.1270D, 2017AJ....154...28B}, there are nearly $3000$ sources identified between $z=4$ and $z=7.54$ to date, yet only $\sim6.5$ per cent of these exhibit radio emission \citep{2017FrASS...4....9P} detected at 1.4~GHz by either the Very Large Array (VLA) Faint Images of the Radio Sky at Twenty-Centimeters\footnote{\url{http://sundog.stsci.edu/}} \citep[FIRST,][]{1995ApJ...450..559B,2015ApJ...801...26H} survey or the  National Radio Astronomy Observatory (NRAO) VLA Sky Survey\footnote{\url{https://www.cv.nrao.edu/nvss/}} \citep[NVSS,][]{1998AJ....115.1693C}. The remaining sources are either outside the footprint of both FIRST and NVSS, or are below the detection limit of the two surveys.

Several studies are available in the literature on the overall radio luminosity function (LF) for AGN  \citep[e.g.,][]{1991ASPC...18..113C,1998MNRAS.300..625W,1999ApJ...511..612G,2014MNRAS.445..955B,2016MNRAS.460....2P,2017ApJ...842...87M,2017A&A...602A...6S}. Many authors found evidence for bimodality in the distribution of the AGN population \citep[e.g.,][]{1980AnA....88L..12S,1989AJ.....98.1195K,1999ApJ...511..612G,2007ApJ...654...99W}, dividing the population to distinct samples of radio-loud and radio-quiet objects. The validity of this dichotomy was recently challenged and found to be the result of observational and mathematical bias effects \citep{2008MNRAS.387..856Z,2012ApJ...754...12M,2012ApJ...759...30B,2013ApJ...768...37C}. 
The weak radio emitters with flux densities below the detection threshold ($\sim1$~mJy for FIRST and $\sim2.5$~mJy for NVSS) might also exhibit radio jets at parsec scales \citep[e.g.,][]{2017ApJ...835L..20W}, but generally remain silent and hidden in flux density limited surveys due to their low radio power and large distance. Faint objects with radio flux densities below the FIRST detection limit can either host radio-quiet AGN, or harbour radio-loud quasars that reside at higher redshifts.

The lurking low- and modest-power members of the high-redshift AGN (hAGN) population can be uncovered by going below the survey threshold and estimating their typical flux densities by means of radio image stacking \citep[e.g.,][]{2007ApJ...654...99W}. Numerous works with stacking analysis are available in the literature on various samples of radio-weak quasars. Analysis of stacked FIRST images at $\sim 8000$ radio-quiet quasar positions from the 2dF QSO redshift survey at medium redshifts ($z\lesssim2.3$) resulted in median flux density levels between $20$ and $40~\mu$Jy  \citep{2005MNRAS.360..453W}. \citet{2008AJ....136.1097H} also found 10s of $\mu$Jy flux densities for a mixed set of Sloan Digital Sky Survey (SDSS) and Luminous Red Galaxy samples. Stacking of extremely red quasars at $2\leq z\leq 4$ using VLA maps at $1.4$ and $6.2$~GHz yielded similar values \citep{2018MNRAS.477..830H}. 

There is a debate on the nature of physical processes responsible for the radio emission at the faint end ($P_\mathrm{1.4GHz}\leq10^{22.5-23}$~W~Hz$^{-1}$) of the continuous radio LF, with possible contribution of AGN jets \citep{1997ARAnA..35..607Z,2018ApJ...869..117R}, AGN winds  \citep{2010ApJ...711..125J,2014MNRAS.442..784Z,2016MNRAS.455.4191Z,2018ApJ...869..117R}, and enhanced star formation or starburst events in the host galaxy \citep{2004MNRAS.352..399J,2013ApJ...768...37C,2013ApJ...778...94R,2018MNRAS.477..830H}.  The origin of radio emission in radio-quiet AGN was also hypothesized to be driven by magnetically heated accreation disk coronae \citep{2008MNRAS.390..847L,2019MNRAS.482.5513L}.
Upper limits on star formation rate (SFR) determined in stacking studies vary between a few to $10$~M$_\odot$~yr$^{-1}$ \citep[e.g.,][]{2007AJ....134..457D,2008AJ....136.1097H}. In other works, low-luminosity AGN jets and  circumnuclear star-forming regions in the host galaxies together are considered responsible for the radio emission \citep[e.g.,][]{2011ApJ...730...61K,2011ApJ...742...45P}.
It was also found that the radio LF usually peaks at higher redshifts for radio-loud sources than for radio-quiet AGN \citep{1999ApJ...511..612G,2017ApJ...842...87M}. 
A multi-wavelength study of  a radio-selected Cosmic Evolution Survey (COSMOS) field sample ($S_\mathrm{1.4GHz}\geq37~\mu$Jy) at $z\leq6$ suggested star formation as the dominant process for the sub-mJy radio population, explaining the dichotomy with AGN activity modes and connection to AGN--host galaxy feedback \citep{2017AnA...602A...3D}. Another study using a sample with $S_\mathrm{1.4GHz}\geq11.5~\mu$Jy applied spectral energy distribution fitting and led to similar conclusions: the peak of the radio LF is at $z\sim2$ \citep{2018AnA...620A.192C}. The radio loudness dichotomy was addressed also at low radio frequencies \citep[$120-168$~MHz,][]{2019AnA...622A..11G}, concluding that low-power quasars are dominated by star formation.

In this paper, we utilised $2229$ empty-field FIRST radio maps centred at positions of hAGN (quasars) identified in the optical or near-infrared. We analysed them with mean and median stacking methods to uncover the underlying radio quasar population, and to examine the possibility of the radio emission originating from either AGN activity or star formation. The aspects of sample selection are detailed in Section~\ref{sec:sample}. We describe the stacking procedure in Section~\ref{sec:stacking}. Results and derived properties are discussed in Section~\ref{sec:discussion}. We summarise our findings and conclude the paper in Section~\ref{sec:summary}.

Throughout this work, we assumed a standard $\Lambda$CDM cosmological model with $\Omega_\mathrm{m}=0.3, \Omega_{\Lambda}=0.7$, and $H_0=70$~km~s$^{-1}$~Mpc$^{-1}$ for calculations.

\section{Sample selection and data}\label{sec:sample}
\begin{figure}
    \centering
    \includegraphics[width=\linewidth]{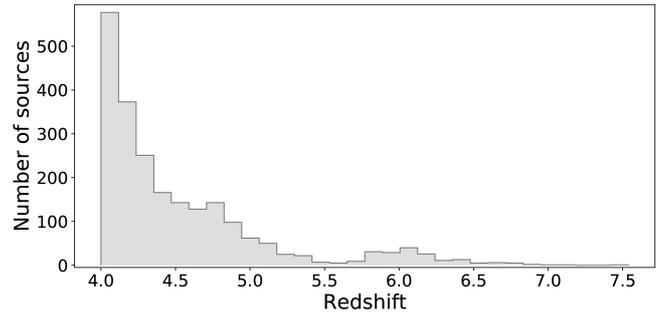}
    \caption{Redshift distribution of the $2229$ high-redshift active galactic nuclei used for the stacking analysis.}
    \label{fig:histogram}
\end{figure}

The sample for the stacking analysis was defined by using the high-redshift ($z \geq 4$) AGN catalogue by \citet{2017FrASS...4....9P}{\footnote{The regularly updated version of the catalogue can be accessed at \url{http://astro.elte.hu/~perger/catalog.html}}}. We selected optically identified AGN from the catalogue with positions falling into the FIRST survey footprint. The flux density of the objects selected is below the detection limit of the survey (typically $\sim1$~mJy). At the time of the analysis, the total number of such AGN was $2232$. Intensity  fluctuations (rms noise) at the optical AGN positions in the FIRST images vary in the range from tens of $\mu$Jy to $\lesssim 1$~mJy. To test an underlying high-redshift radio AGN population, we stacked  FIRST radio maps centred at these positions. The images in Flexible Image Transport System (FITS) format \citep{1981AnAS...44..363W} were obtained from the FIRST image cutout service\footnote{\url{https://third.ucllnl.org/cgi-bin/firstcutout}}. We downloaded images of $4\farcm5\times4\farcm5$ size for each individual AGN position. After excluding the largely incomplete or entirely ragged maps, we obtained $2229$ images for the stacking process. The redshift distribution of the stacked sample is shown in Fig.~\ref{fig:histogram}, and the first 5 entries of the list of $2229$ AGN are given in Table~\ref{tab:sample}. The full list is available in the supplementary material.

\begin{table*}
    \centering\caption{The first 5 entries of the AGN sample used for the stacking analysis.}
    \begin{tabular}{lllll}
         \hline\hline
        Name                        &   R.A.  (h~m~s)                 &   Dec  ($\degr~'~''$)                   &   Redshift & Reference\\
    \hline
SDSS~J000046.69+010951.2	&	$00~00~46.69$	&	$+01~09~51.24$	&	4.25	&	\citet{2012AnA...548A..66P}	\\
SDSS~J000124.23+111212.6	&	$00~01~24.23$	&	$+11~12~12.69$	&	4.30	&	\citet{2017AnA...597A..79P}	\\
SDSS~J000404.71+000039.0	&	$00~04~04.71$	&	$+00~00~39.08$	&	4.31	&	\citet{2012AnA...548A..66P}	\\
SDSS~J000457.11$-$000538.7	&	$00~04~57.11$	&	$-00~05~38.78$	&	4.05	&	\citet{2012AnA...548A..66P}	\\
SDSS~J000527.14+025813.2	&	$00~05~27.15$	&	$+02~58~13.29$	&	4.11	&	\citet{2012AnA...548A..66P}	\\
    \hline
    \end{tabular}\\
   {\it Notes.} Column 1 -- object name, Column 2 -- right ascension, Column 3 -- declination, Column 4 -- spectroscopic redshift, Column 5 -- literature reference of discovery. The full table of $2229$ AGN is available in the electronic version of the journal. 
    \label{tab:sample}
\end{table*}

\section{Stacking analysis}\label{sec:stacking}


\subsection{Stacking of catalogue sources}

The sample of $2229$ objects were binned based on the redshift of each source into four subsamples with approximately equal number of images in each bin. There are $554$, $559$, $559$, and $557$ objects in the four bins with redshift boundaries of $4.0\leq z_1<4.1,4.1\leq z_2<4.3,4.3\leq z_3<4.7,$ and $4.7\leq z_4<7.6$, respectively.

We performed the stacking in both mean and median procedures. Both the mean  \citep[e.g.][]{2007ApJ...654...99W,2007AJ....134..457D,2008AJ....136.1097H} and median \citep[e.g.][]{2007ApJ...654...99W,2007AJ....134..457D,2008AJ....136.1097H,2018MNRAS.477..830H,	2018ApJ...869..117R} stacking methods are commonly used in the literature. Mean stacking determines the arithmetic average of intensities in the image pixels within the sample. However, this method is very sensitive to outlying values in the sample, as well as to the threshold set to avoid the contaminating point sources. In turn, median stacking deals with the median of intensity values corresponding to the image pixels in the sample.

To evaluate the applicability of maps resulted from both methods, we calculated root-mean-square (rms) noise, maximum of the intensity, and signal-to-noise ratio (SNR) for both the mean and median stacked images in each bin and for the full sample.
Typical rms noise values in the separate redshift bins span the range $10-20~\mu$Jy~beam$^{-1}$ for mean stacked images, while median stacking resulted in $\sim 7\mu$Jy~beam$^{-1}$ noise levels. The full-sample maps have $8~\mu$Jy~beam$^{-1}$ and $3~\mu$Jy~beam$^{-1}$ rms noise for the mean and median methods, respectively.

\begin{table}
\centering
\caption{Properties of median-stacked maps.}\label{tab:data}
\begin{tabular}{cccc}
\hline\hline
Bin & rms [$\mu$Jy~beam$^{-1}$] & $I_\mathrm{max} $ [$\mu$Jy~beam$^{-1}$] & SNR \\
\hline
1	&		7		&		27		& 	4		\\
2	&		7		&		52		& 	7		\\
3	&		7		&		38		& 	5		\\
4	&		7		&		30		& 	4		\\
All	&		3		&		35		& 	11		\\

\hline
\end{tabular}\\
{\it Notes.} Column 1 -- redshift bin, Column 2 -- image noise, Column 3 -- maximum intensity, Column 4 -- signal-to-noise ratio
\end{table}

Despite the high SNR values (up to $50$) found in each mean-stacked map, the method provides insignificant results in the search of a hidden central source. The maps are contaminated and dominated by peaks at random off-centre locations, caused by strong intensity peaks from the stacked individual maps. The presence of off-centre sources leads to an increase of the rms image noise in the field compared to median stacking. Therefore mean-stacked maps are not considered in the further study and calculations. 

\begin{figure*}

\centering
    \subfloat[Median-stacked maps for the four redshift bins.]{\includegraphics[width=0.85\linewidth]{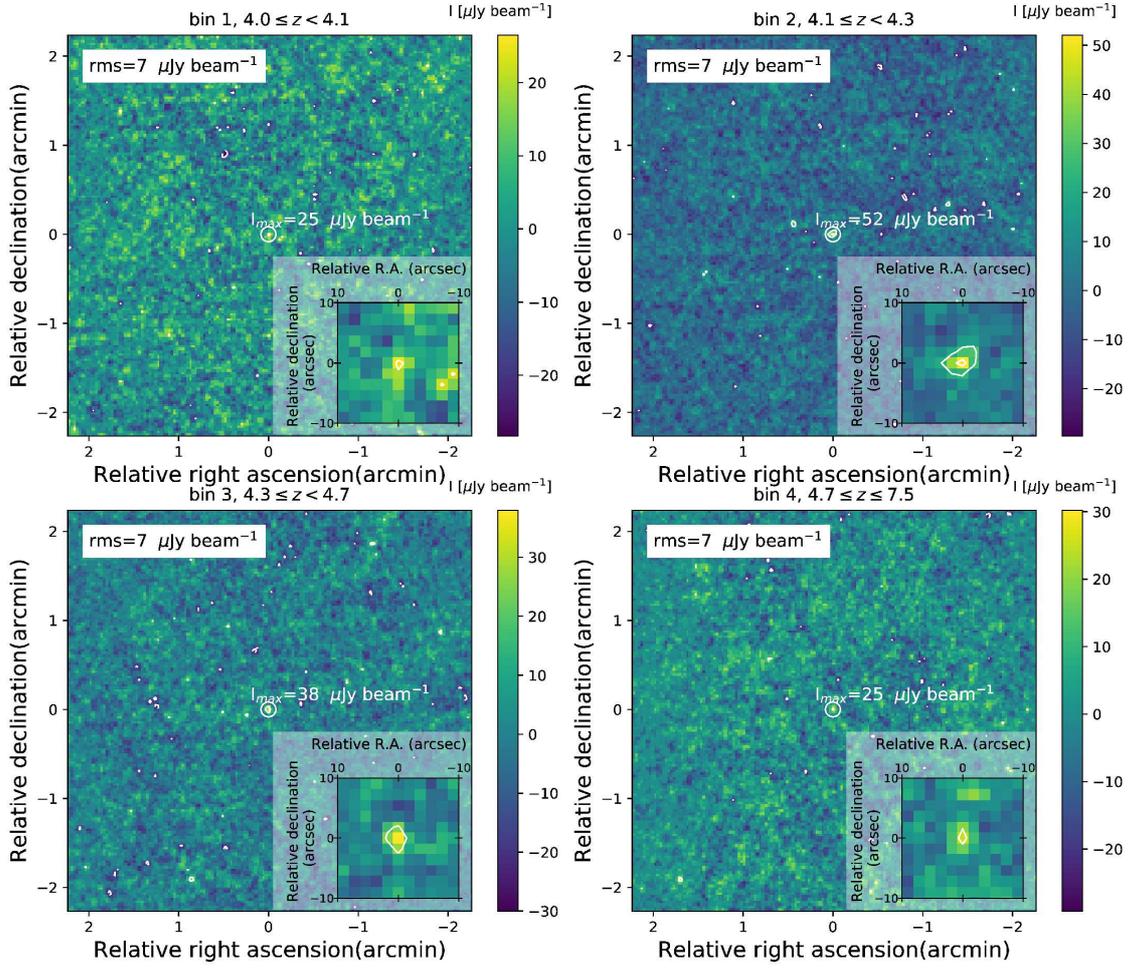}}\\
    \subfloat[Median-stacked map derived from all images in the sample. The inner $30\arcsec \times 30\arcsec$ area is shown.\label{fig:bin_all}]{\includegraphics[width=0.85\linewidth]{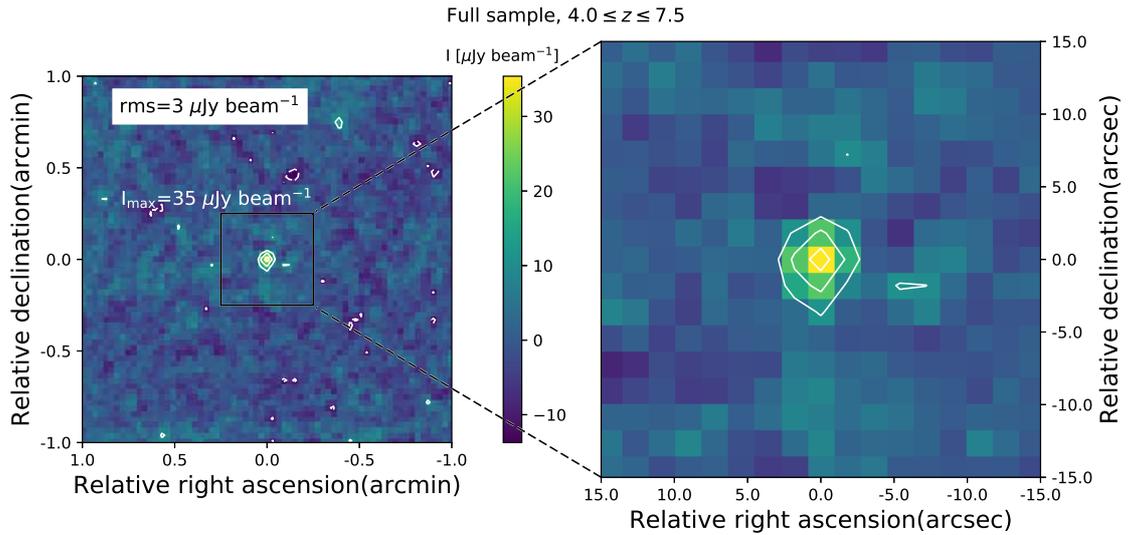}}
\caption{The central $2\arcmin \times 2\arcmin$ area of the median-stacked FIRST image cutouts of $z\geq 4$ optically identified radio-non-detected AGN. For the binned data, the inner $20''\times20''$ area is also shown in the insets. Maximum intensities in the central 5 arcsec radius circle areas are indicated. The first contours are drawn at $\pm3$ times the rms noise level. Additional contours are at $6$ and $9$ times the rms. }\label{fig:bins}
\end{figure*}

\begin{figure}
    \centering
    \includegraphics[width=\linewidth]{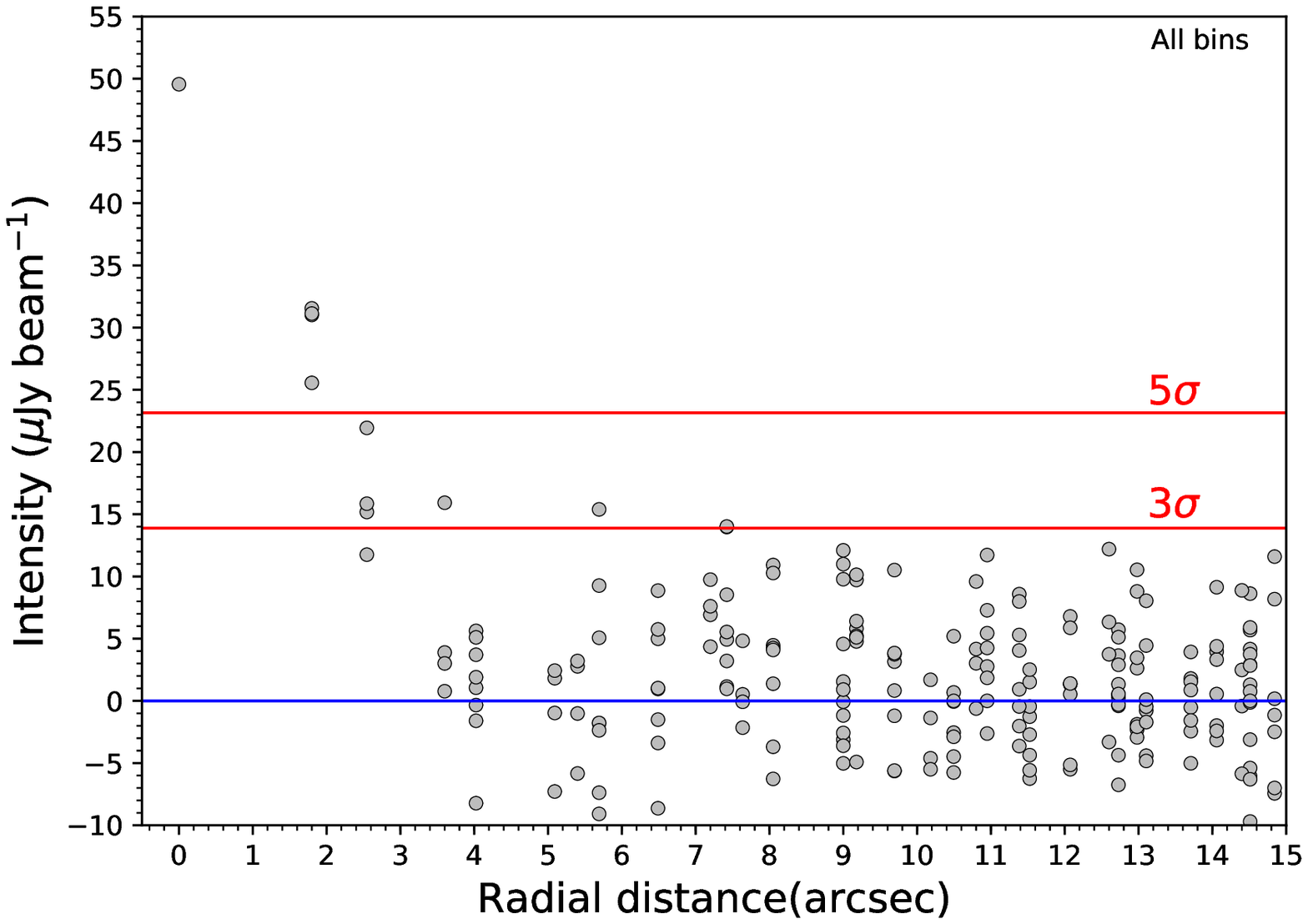}
    \caption{Pixel-by-pixel radial plot of the full-sample median-stacked intensity values for the inner $15''$ radius area. 1 pixel equals $1.8''$. For the data points, the unbiased values are shown, after applying the bias correction factor $1.4$ \citep{2007ApJ...654...99W}.}
    \label{fig:radial}
\end{figure}

Median maps on the contrary are not sensitive to occasional bright off-centre sources. They show a nearly uniform noise level in each subsample ($7~\mu$Jy~beam$^{-1}$), revealing a protruding radio peak at the image centre for the entire sample with SNR exceeding 10. Intensity maxima for the median-stacked images are found to be $27, 52, 38$ and $30~\mu$Jy~beam$^{-1}$ for the redshift-binned data, and $35~\mu$Jy~beam$^{-1}$ for the full sample. Radio maps obtained by median stacking are shown in Fig.~\ref{fig:bins}, for which the calculated image properties are listed in Table~\ref{tab:data}, while the radial profile of the full-sample median-stacked image is illustrated in Fig.~\ref{fig:radial}.

The image rms noise reached by stacking is expected to decrease with respect to the single-image rms by $\sqrt N$, where $N$ is the number of stacked images. The original FIRST image noise levels are $\approx 0.15$~mJy. For the sample of $N=2229$ objects, this predicts the rms of $150~\mu$Jy~beam$^{-1}$ divided by $\sqrt{2229}$, i.e. $\approx 3~\mu$Jy~beam$^{-1}$ to be reached with stacking. Indeed, this is equal to the actual rms values obtained for the full-sample median-stacked images.

\subsection{Stacking of fake sources and the image noise}
To check the reliability of our results, we created four samples of the same size, with arbitrary source coordinates in the manner of adding or subtracting $1\degr$ to/from either or both the right ascensions and declinations of objects in the  the original sample. Binning was based on the redshifts of AGN in the original (true) positions.

We did not find any radio emission in either of the four samples of fake sources, neither with mean, nor with median stacking methods. The mean images show outlier sources outside the central pixels, similarly to the results obtained for the true sample. The rms noise levels of the binned subsample maps are $10-20~\mu$Jy~beam$^{-1}$ and $\sim7~\mu$Jy~beam$^{-1}$ for the mean and median stacking, respectively. Stacking of all images resulted in rms values comparable to that of the real sample and consistent with the expected values from the stacking procedure, $8~\mu$Jy~beam$^{-1}$ for mean and $3~\mu$Jy~beam$^{-1}$ for median maps. The SNR values for the mean images vary from $16$ to $50$ in the separate redshift bins, and are below $\approx 30$ when the full fake samples were included in mean stacking. The image peaks are caused by occasional bright off-centre sources appearing in some fields, similarly to the real sample. For median stacking, the typical values in the fabricated samples and in their subsets are in the range of $4\lesssim \mathrm{SNR}\lesssim5$, and indicate no significant detection at the central location, as expected. Due to the similar SNR values in the samples of fake sources, the binned data were not considered for further analysis, only the full sample.

\subsection{Model fit to the stacked data}

For further analysis, we fitted a circular Gaussian model component to the central region of the full-sample median-stacked image, to characterise the brightness distribution using the {\sc imfit} task in the U.S. National Radio Astronomy Observatory (NRAO) Astronomical Image Processing System\footnote{http://www.aips.nrao.edu/index.shtml} \citep[{\sc aips},][]{2003ASSL..285..109G}. This resulted in an unresolved point source. By multiplying the fitted model component flux density with the bias correction factor of $1.4$ suggested by \citet{2007ApJ...654...99W}, we obtained an unbiased value of $S_\mathrm{1.4GHz}=52\pm1\,\mu{\mathrm Jy}$.


\subsection{Characteristic 1.4-GHz radio power }

To derive a characteristic value of the monochromatic rest-frame 1.4-GHz radio power for the hAGN population, we co-added the individual FIRST image cutouts around the positions of the full sample of $2229$ objects that are individually not detected in the FIRST survey. Motivated by the model-fitting results of the median stacking, we assumed that the radio emission originates from point sources unresolved in FIRST. Thus the flux density was derived from the cumulative brightness value of the central pixel for the following analysis. We consider that the co-added flux density of the $2229$ hAGN is the sum of the flux densities of the individual sources, and that all hAGN included in the stacking process have the same radio power at $1.4$~GHz, $P_\mathrm{1.4GHz}$. This latter assumption is obviously not true, but the value derived this way is characteristic to the hAGN sample as a whole. 

We used the following relationship between flux density and power (including $K$-correction):
\begin{equation}
    S_\mathrm{1.4GHz,sum}=\sum_{i=1}^{N}\frac{P_\mathrm{1.4GHz}}{4\pi D_{\mathrm{L}i}^2(1+z_i)^{-\alpha-1}},
\end{equation}
where $S_\mathrm{1.4GHz,sum}=77$~mJy is the cumulative flux density derived from co-adding the $N=2229$ FIRST map centres, $\alpha$ is the radio spectral index (using the convention $S_\nu\propto\nu^\alpha$, where $\nu$ is the frequency), and $D_{\mathrm{L}i}$ and $z_i$ are the luminosity distance and the redshift of the $i$-th individual source, respectively. Since the individual power-law spectral indices are unknown for the sources, we assumed a common value for the whole sample. Various radio spectral indices appropriate for galaxies are found in the literature \citep[e.g.][]{2003ApJ...599..971H,2007AnA...462..525H,2007ApJ...654...99W,2018MNRAS.477..830H}. Therefore we calculated the characteristic power considering several different values in the range $-0.5\leq\alpha\leq-1$, corresponding to steep radio spectra. The estimated values of the $1.4$-GHz power for each spectral index are listed in Col.~2 in Table~\ref{tab:powsfr}.


\subsection{Upper limit on star formation}

Radio emission from galaxies may originate from AGN activity in the nuclear region where synchrotron radiation is produced in powerful relativistic plasma jets driven by accretion onto the central supermassive black hole. On the other hand, synchrotron emission from relativistic electrons and free-free emission from ionized hydrogen regions may be widespread in the AGN host galaxy as a consequence of recent start formation \cite[e.g.][]{1992ARA&A..30..575C}. 

We can assume that the radio emission in our stacked hAGN sample originates solely from star formation taking place around the central regions of the host galaxy, rather than from AGN activity. This way we can calculate an upper limit on the star formation rate (SFR) using the estimated characteristic $1.4$-GHz power value, applying various correlations between the radio power and SFR from the literature \citep{2003ApJ...586..794B,2006ApJ...643..173S,2017MNRAS.466.4917D,2019MNRAS.482..560M}.  The SFR values obtained are listed in Cols.~3--6 in Table~\ref{tab:powsfr}. The upper limits on the SFR span about an order of magnitude, depending on the radio spectral index assumed and the correlation used. But we can conclude that the typical SFR upper limits are in the order of $10^3$~M$_\odot$~yr$^{-1}$.

\begin{table}
    \centering
    \caption{Estimated characteristic 1.4-GHz radio powers and star formation rates for the radio sources associated with the central pixels of the co-added $2229$ FIRST maps.}
    \begin{tabular}{cccccc}
    \hline\hline
        \multirow{2}{*}{$\alpha$}  &\multirow{2}{*}{$P_\mathrm{1.4GHz}$ [$10^{24}$~W~Hz$^{-1}$]}	&\multicolumn{4}{c}{SFR [M$_\odot$~yr$^{-1}$]}	 \\
            	&	        	&	(1)	    &	(2)	    &	(3)	    &	(4) 	\\ \hline
        $-0.5$  &  $2.9$   &  $1600$   & $1800$   & $370$   &  $1700$  \\
        $-0.6$  &  $3.4$   &  $1900$   & $2100$   & $420$   &  $2000$  \\
        $-0.7$  &  $4.1$   &  $2300$   & $2500$   & $480$   &  $2400$  \\
        $-0.8$  &  $4.8$   &  $2700$   & $3000$   & $540$   &  $2800$  \\
        $-0.9$  &  $5.7$   &  $3200$   & $3500$   & $610$   &  $3400$  \\
        $-1 $   &  $6.8$   &  $3700$   & $4200$   & $700$   &  $4000$ \\

    \hline
    \end{tabular}\label{tab:powsfr}\\
    {\it Notes.} Column 1 -- assumed spectral index, Column 2 -- 1.4-GHz radio power, Column 3--6 -- star formation rates calculated using various relationships by (1)~\citet{2003ApJ...586..794B}, (2)~\citet{2006ApJ...643..173S}, (3)~\citet{2017MNRAS.466.4917D}, and (4)~\citet{2019MNRAS.482..560M}
\end{table}


\section{Discussion}\label{sec:discussion}

\subsection{Flux densities}

The flux density of $52~\mu$Jy found in our median stacking analysis is comparable but somewhat higher than the values obtained in previous stacking works \citep[e.g.][]{2005MNRAS.360..453W,2007ApJ...654...99W,2008AJ....136.1097H,2018MNRAS.477..830H}. Considering that our sources are located at higher redshifts, this suggests that the hAGN are generally intrinsically more powerful than the sources studied in those samples.  

For comparison, the ultraluminous quasar SDSS~J010013.02+280225.8 at $z=6.3$ --  not included in our stacking analysis due to its position outside the FIRST survey footprint -- was detected with the VLA at a higher frequency, 3~GHz ($0.65''$ resolution and $\sim3~\mu$Jy~beam$^{-1}$ sensitivity), with a flux density of $\sim100~\mu$Jy \citep{2016ApJ...830...53W}. Assuming a spectral index of $-0.7$, this scales to $\sim170~\mu$Jy at $1.4$~GHz. Further high-resolution observations performed with the Very Long Baseline Array (VLBA) at $1.5$~GHz revealed a radio structure partially resolved at $\sim 10$ milli-arcsecond (mas) level, with $\sim90~\mu$Jy flux density \citep{2017ApJ...835L..20W}. The detection of a mas-scale radio structure in this individual source raises the possibility that the FIRST-undetected sources in our stacked sample could also be targeted with sensitive very long baseline interferometry (VLBI) observations in the future, in a hope of revealing weak compact AGN-related radio emission. It also suggests that the radio emission in at least some of the sources in our $2229$-element sample does have synchrotron AGN jet contribution and is not associated with star formation only. 

\subsection{Radio power and AGN radio emission}

All estimated values for the $1.4$-GHz radio power are $P_\mathrm{1.4GHz}\sim10^{24}$~W~Hz$^{-1}$ (Table~\ref{tab:powsfr}) in this study. If the radio emission in these objects would originate solely from AGN activity, the FIRST-undetected hAGN would be among the radio AGN population ($P_\mathrm{1.4GHz}\geq2\times10^{22.5}$~W~Hz$^{-1}$), since powers exceed the threshold found in empirical studies between supernova-related and AGN activity  \citep[$P_\mathrm{1.4GHz}=2\times10^{21}$~W~Hz$^{-1}$,][]{2000ApJ...530..704K, 2011AnA...526A..74M}. 
The estimated radio powers also surpass values measured for prominent starburst galaxies, e.g. Arp\,220, Arp\,229A, and Mrk\,273 \citep[$P_\mathrm{1.4GHz}=2-4.5\times10^{22}$~W~Hz$^{-1}$,][]{2012MNRAS.423.1325A}. 

Monochromatic powers found in this work are 2 to 5 orders of magnitude higher than those found for low-power sources at low redshifts ($z<0.3$), e.g. Lyman-break analogues (LBAs) including an LBA with dominant star formation (J0150+1308), an AGN (J1029+4829) and an AGN--SFR composite source (J0921+4509) \citep{2012MNRAS.423.1325A}. Also, low-redshift AGN SDSS\,J1155+1507,  SDSS\,J2104$-$0009, SDSS\,J2304$-$0933 \citep{2016ApJ...826..106G}, and  NGC\,3147 \citep{2004ApJ...603...42A} exhibit radio powers in the range of $5\times10^{21}$~W~Hz$^{-1}$ to a few times $10^{22}$~W~Hz$^{-1}$, or even lower powers between $10^{19}$~W~Hz$^{-1}$ to a few times $10^{20}$~W~Hz$^{-1}$ in the AGN Henize\,2-10 \citep{2012ApJ...750L..24R}, NGC\,4203, NGC\,4535 \citep{2002ApJ...581..925U,2004ApJ...603...42A}, NGC\,864, and NGC\,4123 \citep{2002ApJ...581..925U}. It was discussed in e.g. \citet[]{2005MNRAS.362...25B,2008ASPC..399..413S,2013MNRAS.433.2647S} and \citet{2016MNRAS.455.2731R} that the most luminous AGN have radio powers in the range from $10^{24}$~W~Hz$^{-1}$ to a few times $10^{26}$~W~Hz$^{-1}$, which is comparable with our results. These studies of low-redshift AGN found that host galaxies with $1.4$-GHz powers above $10^{23}$~W~Hz$^{-1}$ usually host radio-loud AGN in their centres. This outlines that AGN activity represents a significant contribution to the 1.4~GHz radio flux densities in the stacked sample.

Deep radio surveys at high redshift, such as the Extended Chandra Deep Field-South \citep[E-CDFS,][]{2008ApJS..179..114M,2013ApJS..205...13M} survey and the VLA-COSMOS 3 GHz Large Project \citep{2017A&A...602A...1S} can provide additional information about the most distant galaxy centres. 
Our estimated radio powers ($\sim 10^{24}$~WHz$^{-1}$) are supported by the results achieved by e.g. \citet{2017AnA...602A...3D}, \citet{2017A&A...602A...2S}, and \citet{2017A&A...602A...6S}. In a multiwavelength analysis of $z\leq6$ radio sources in the COSMOS field at 3~GHz, \citet{2017AnA...602A...3D} divided the sample into three populations: radio-quiet (RQ) and radio-loud (RL) AGN, and star-forming galaxies (SFGs). The majority of RQ AGN hosts show enhanced star formation. Similar results were obtained by \citet{2017A&A...602A...2S}, revealing that radio sources at $1.4$~GHz with flux densities above $\sim200~\mu$Jy are predominantly AGN, and that with decreasing flux densities the SFGs take over as the dominant population in up to $\sim60$~per cent of the sample.
However, the $1.4$~GHz luminosity functions determined for AGN with radio excess (with respect to the expected contribution of SF to the radio emission) using different procedures by \citet{2017A&A...602A...6S} showed that rest-frame radio powers span the range of $\sim 10^{24}-10^{26}~$WHz$^{-1}$ for AGN at the highest redshifts, when relying solely on observational data, and 
$10^{22}-10^{27}~$WHz$^{-1}$ with the application of evolution models. The fractional distribution found in the COSMOS field studies would imply an RQ--SFG dominance in the sample. Our estimated radio powers suggest that the objects in the stacking analysis are mostly radio-loud AGN.
Another study of radio sources below $\sim100~\mu$Jy based on the E-CDFS survey concluded that besides the dominance of SFGs and the decrase of the RL AGN towards the lowest flux densities, the number of RQ AGN increases \citep{2013MNRAS.436.3759B}. They also determined radio power distributions for all three populations, and found median values of both RQ and RL AGN which coincide with our estimated radio powers, while the power at the peak of the distribution for SFGs is an order of magnitude lower.

\subsection{Star formation rate}

SFR upper limits derived in this paper (Table~\ref{tab:powsfr}) are two to three orders of magnitude higher than SFR values found for individual AGN host galaxies in the literature, e.g.  $0.5-2$~M$_\odot$~yr$^{-1}$ for SDSS\,J2104$-$0009 and  SDSS\,J2304$-$0933 \citep{2016ApJ...826..106G}, $5-8$~M$_\odot$~yr$^{-1}$ for J0150+1308, J0921+4509, and J1029+4829  \citep{2012MNRAS.423.1325A}, and $\sim1$~M$_\odot$~yr$^{-1}$ with H$_\alpha$ star formation rates predicted by \citet{2002ApJ...581..925U}, as well as SFR with upper limits found in previous works using image stacking \citep[$\leq10$~M$_\odot$~yr$^{-1}$, e.g.][]{2007AJ....134..457D,2008AJ....136.1097H}. Even ultraluminous infrared galaxies were found to show star formation with rates varying between a couple of tens to a few hundreds of M$_\odot$~yr$^{-1}$  \citep[e.g.][]{2010ApJ...715..572H}, suggesting that our upper limits appreciably overestimate the contribution of star formation to the 1.4-GHz radio power.

However, turning to high-redshift objects, CO and [C II] line detection and dust emission in individual $z\sim6$ quasars indicate significant star formation in the central few kpc region in some host galaxies, with SFR in the order of $\sim1000$~M$_\odot$~yr$^{-1}$ \citep[e.g.][]{2003AnA...409L..47B,2013ApJ...770...13W,2013ApJ...773...44W,2019ApJ...876...99S}. Note that those AGN with an estimated SFR of a few thousand M$_\odot$~yr$^{-1}$ have not been included in our stacked sample because they are outside the FIRST survey coverage. For some individual objects that are included in our FIRST image stacking, molecular and atomic line observations determined a wide range of SFR values.
These start from a few tens of M$_\odot$~yr$^{-1}$, e.g. $48$~M$_\odot$~yr$^{-1}$ for CFHQS\,J0210$-$0456, $<40$~M$_\odot$~yr$^{-1}$ for CFHQS\,J2329$-$0301 \citep{2013ApJ...770...13W}, and $\sim80$~M$_\odot$~yr$^{-1}$ for CFHQS\,J0055+0146 \citep{2015ApJ...801..123W}.  SFR values of a couple of hundreds M$_\odot$~yr$^{-1}$ were found for VIKING\,J2348$-$3054 and VIKING\,J0109$-$3047 \citep[$\sim700$ and $\sim900$~M$_\odot$~yr$^{-1}$, respectively,][]{2016ApJ...816...37V},  ULAS\,J1120+0641  \citep[$\sim200$~M$_\odot$~yr$^{-1}$,][]{2017ApJ...837..146V}, and SDSS\,J0100+2800 \citep[$\sim650$~M$_\odot$~yr$^{-1}$,][]{2016ApJ...830...53W}. There are also examples of sources with SFR up to thousands of M$_\odot$~yr$^{-1}$ \citep[][]{2004ApJ...615L..17W,2019ApJ...874L..30V}. 

The difference between the SFR upper limits we estimated as characteristic for the stacked hAGN sample ($\lesssim4000$~M$_\odot$~yr$^{-1}$, Table~\ref{tab:powsfr}) and the values measured for individual sources using independent methods is naturally explained if we relax the assumption that the entire radio emission comes from star forming activity in the host galaxies, without AGN contribution. Moreover, we calculated with a single characteristic radio power for all sources. In reality, the physical conditions in the individual objects are clearly more complex than just assuming a constant power, and one or the other process solely responsible for the radio emission. In fact, radio emission originating from both AGN and star formation must be present in most objects, in various proportions. 

In a hope to refine the results, one may take the AGN radio luminosity function (LF) into account, up to the highest redshifts \citep[e.g.][]{2017A&A...602A...6S, 2018AnA...620A.192C}. Improvements in determining the radio LF at the earliest cosmological epochs are expected from the currently ongoing $2-4$~GHz VLA Sky Survey \citep[VLASS,][]{2015fers.confE...6M}. It will have higher sensitivity ($\sim70~\mu$Jy for the combined 3-epoch observations) and angular resolution ($2\farcs5$) compared to the FIRST survey. Also, repeating a stacking analysis similar to the one reported here, the deeper VLASS radio maps would allow for determining redshift-dependent properties of the FIRST-undetected AGN. Our study with binned sub-samples did not provide conclusive results because of the insufficient SNR in the stacked images.

\subsection{Origin of the sub-mJy radio emission}
Ascertaining the role of SF and AGN contribution to the  sub-mJy radio emission could be aided by the application of methods independent from the FIRST observations.
Based on mid-infrared (MIR) polycyclic aromatic hydrocarbon (PAH) detections in low-redshift galaxies, PAH emission lines can be found very close to the galaxy core, at $\sim1-20$~kpc distances \citep[e.g.,][]{2007ApJ...669..841S,2019ApJ...871..190M}. Since star formation indicators are found this close to the core, and our stacking analysis showed that the fitted Gaussian model component is unresolved in FIRST (corresponding to $26-35$~kpc linear size depending on the redshift), the angular resolution provided by FIRST is insufficient to distinguish between star formation and AGN related emission. 
Observations with the upcoming Square Kilometre Array (SKA), in cooperation with globally distributed VLBI arrays could provide an adequate mas or sub-mas resolution \citep{2012PASA...29...42G,2015aska.confE.143P} and the thermal sensitivity of a few $\mu$Jy \citep{2015aska.confE.143P} sufficient for direct detection of compact AGN-related radio emission in the most powerful members of the radio-hAGN population.

The importance and contribution of SF to the radio emission can also be tackled using MIR observations. Considering the correlation between the MIR and $1.4$-GHz radio powers, and applying the fit parameters in \citet{2005ApJ...632L..79W}, we calculated the $8~\mu$m and $24~\mu$m powers that correspond to the characteristic $P_\mathrm{1.4GHz}$ values found in our study. Consistency of the MIR--radio correlation is considered invariant for 5 orders of magnitude of flux densities up to $z\sim3.5$  \citep{2001ApJ...554..803Y,2008MNRAS.386..953I} and was found reliable even at $\mu$Jy levels \citep{2009MNRAS.394..105G}. Characteristic MIR flux densities were estimated for our hAGN sample, based on the characteristic power derived from the co-added 1.4-GHz FIRST radio maps. Values of $10-15$~mJy  and $30-50$~mJy were found for the $8~\mu$m and $24~\mu$m flux densities, with mean values $14~$mJy and $45~$mJy, respectively. To estimate the level of AGN contamination contributing to the calculated values of SFR, we determined $q_{24}$ values using FIRST upper limits and  $24~\mu$m flux densities derived using $22~\mu$m emission measured by the Wide-field Infrared Survey Explorer \citep[WISE,][]{2010AJ....140.1868W} from the AllWISE catalog\footnote{http://wise2.ipac.caltech.edu/docs/release/allwise/} \citep{2014yCat.2328....0C}. We followed the analysis described by e.g. \citet{2013MNRAS.436.3759B}. On the one hand, all $124$ sources detected by WISE are above the theoretical limit separating RL AGN from RQ AGN and SFGs, so no further constraints could be obtained. On the other hand, the $24~\mu$m flux densities are in the range of $\sim2-11~$mJy with a mean value of $4~$mJy, which is an order of magnitude lower than the values calculated from the MIR--radio correlation, assuming all radio emission is SF-related. This indicates radio excess for the stacked objects (at least those with WISE detection), implying that the radio emission is AGN-related.

\section{Summary}\label{sec:summary}

We applied median stacking on 1.4-GHz VLA FIRST survey image cutouts centred on $2229$ optically identified but individually radio-undetected quasar positions. These objects populate the redshift range $4\leq z < 7.6$. Stacking of the full sample resulted in an unresolved point source with a bias-corrected flux density $52$~$\mu$Jy. Co-adding the radio map central pixels revealed a moderately radio-loud AGN population, with a characteristic 1.4-GHz radio power $P_\mathrm{1.4GHz}\sim10^{24}$~W~Hz$^{-1}$. Under the simplifying assumption that the entire radio emission in the sample is produced by star forming activity in the quasar host galaxies, we obtained upper limits of the star formation rate in the order of a few $1000$~M$_\odot$~yr$^{-1}$. Based on literature data on individual AGN, we argue that the source of radio emission in the sample is rather a mixture of star formation and AGN-related activity. The spatial resolution of FIRST images is not sufficient to distinguish between the different mechanisms responsible for the radio emission. Future measurements with VLBI and SKA-VLBI could help determining the relative importance of the two emission types in individual objects. Stacking studies similar to the one presented here will benefit from the improved sensitivity and angular resolution of the ongoing VLASS whose radio images could provide data sufficient for determining redshift-dependent properties of high-redshift radio quasars.

\section*{Acknowledgements}
We are grateful for the constructive comments by the anonymous referee that led to an improved discussion of our results.
This publication makes use of data products from the Wide-field Infrared Survey Explorer, which is a joint project of the University of California, Los Angeles, and the Jet Propulsion Laboratory/California Institute of Technology, funded by the National Aeronautics and Space Administration.
K\'{E}G was supported by the J\'{a}nos Bolyai Research Scholarship of the Hungarian Academy of Sciences, and by the Ministry of Human Capacities within the framework of the \'UNKP (New National Excellence Program).


\bsp	
\label{lastpage}

\begin{thebibliography}{}%

\bibitem[\protect\citeauthoryear{Alexandroff et al.}{2012}]{2012MNRAS.423.1325A} Alexandroff R., et al., 2012, \mnras, 423, 1325 
\bibitem[\protect\citeauthoryear{Anderson, Ulvestad \& Ho}{2004}]{2004ApJ...603...42A} Anderson J.~M., Ulvestad J.~S., Ho L.~C., 2004, \apj, 603, 42 
\bibitem[\protect\citeauthoryear{Ba{\~n}ados et al.}{2018}]{2018Natur.553..473B} Ba{\~n}ados E., et al., 2018, \nat, 553, 473 
\bibitem[\protect\citeauthoryear{Balokovi{\'c} et al.}{2012}]{2012ApJ...759...30B} Balokovi{\'c} M., Smol{\v c}i{\'c} V., Ivezi{\'c} {\v Z}., Zamorani G., Schinnerer E., Kelly B.~C., 2012, \apj, 759, 30 
\bibitem[\protect\citeauthoryear{Becker, White \& Helfand}{1995}]{1995ApJ...450..559B} Becker R.~H., White R.~L., Helfand D.~J., 1995, \apj, 450, 559 
\bibitem[\protect\citeauthoryear{Bell}{2003}]{2003ApJ...586..794B} Bell E.~F., 2003, \apj, 586, 794 
\bibitem[\protect\citeauthoryear{Bertoldi et al.}{2003}]{2003AnA...409L..47B} Bertoldi F., et al., 2003, \aap, 409, L47 
\bibitem[\protect\citeauthoryear{Best et al.}{2005}]{2005MNRAS.362...25B} Best P.~N., Kauffmann G., Heckman T.~M., Brinchmann J., Charlot S., Ivezi{\'c} {\v{Z}}., White S.~D.~M., 2005, \mnras, 362, 25
\bibitem[\protect\citeauthoryear{Best et al.}{2014}]{2014MNRAS.445..955B} Best P.~N., Ker L.~M., Simpson C., Rigby E.~E., Sabater J., 2014, \mnras, 445, 955
\bibitem[\protect\citeauthoryear{Blanton et al.}{2017}]{2017AJ....154...28B} Blanton M.~R., et al., 2017, \aj, 154, 28 
\bibitem[\protect\citeauthoryear{Bonzini et al.}{2013}]{2013MNRAS.436.3759B} Bonzini M., et al., 2013, \mnras, 436, 3759
\bibitem[\protect\citeauthoryear{Ceraj et al.}{2018}]{2018AnA...620A.192C} Ceraj L., et al., 2018, \aap, 620, A192 
\bibitem[\protect\citeauthoryear{Chambers et al.}{2016}]{2016arXiv161205560C} Chambers K.~C., et al., 2016, arXiv:1612.05560 
\bibitem[\protect\citeauthoryear{Condon}{1991}]{1991ASPC...18..113C} Condon J.~J., 1991, \aspc, 18, 113 
\bibitem[\protect\citeauthoryear{Condon}{1992}]{1992ARA&A..30..575C} Condon J.~J., 1992, \araa, 30, 575 
\bibitem[\protect\citeauthoryear{Condon et al.}{1998}]{1998AJ....115.1693C} Condon J.~J., Cotton W.~D., Greisen E.~W., Yin Q.~F., Perley R.~A., Taylor G.~B., Broderick J.~J., 1998, \aj, 115, 1693 
\bibitem[\protect\citeauthoryear{Condon et al.}{2013}]{2013ApJ...768...37C} Condon J.~J., Kellermann K.~I., Kimball A.~E., Ivezi{\'c} {\v Z}., Perley R.~A., 2013, \apj, 768, 37 
\bibitem[\protect\citeauthoryear{Cutri et al.}{2014}]{2014yCat.2328....0C} Cutri R.~M., et al., 2014, yCat, II/328
\bibitem[\protect\citeauthoryear{Dark Energy Survey Collaboration et al.}{2016}]{2016MNRAS.460.1270D} Dark Energy Survey Collaboration, et al., 2016, \mnras, 460, 1270 
\bibitem[\protect\citeauthoryear{Davies et al.}{2017}]{2017MNRAS.466.4917D} Davies R.~I., et al., 2017, \mnras, 466, 4917 
\bibitem[\protect\citeauthoryear{de Gasperin et al.}{2011}]{2011MNRAS.415.2910D} de Gasperin F., Merloni A., Sell P., Best P., Heinz S., Kauffmann G., 2011, \mnras, 415, 2910 
\bibitem[\protect\citeauthoryear{de Vries et al.}{2007}]{2007AJ....134..457D} de Vries W.~H., Hodge J.~A., Becker R.~H., White R.~L., Helfand D.~J., 2007, \aj, 134, 457 
\bibitem[\protect\citeauthoryear{Delvecchio et al.}{2017}]{2017AnA...602A...3D} Delvecchio I., et al., 2017, \aap, 602, A3 
\bibitem[\protect\citeauthoryear{Djorgovski et al.}{2001}]{2001ASPC..225...52D} Djorgovski S.~G., Mahabal A.~A., Brunner R.~J., Gal R.~R., Castro S., de Carvalho R.~R., Odewahn S.~C., 2001, \aspc, 225, 52 
\bibitem[\protect\citeauthoryear{Eisenstein et al.}{2011}]{2011AJ....142...72E} Eisenstein D.~J., et al., 2011, \aj 142, 72 
\bibitem[\protect\citeauthoryear{Filho et al.}{2004}]{2004AnA...418..429F} Filho M.~E., Fraternali F., Markoff S., Nagar N.~M., Barthel P.~D., Ho L.~C., Yuan F., 2004, \aap, 418, 429 
\bibitem[\protect\citeauthoryear{Gab{\'a}nyi et al.}{2016}]{2016ApJ...826..106G} Gab{\'a}nyi K.~{\'E}., An T., Frey S., Komossa S., Paragi Z., Hong X.-Y., Shen Z.-Q., 2016, \apj, 826, 106 
\bibitem[\protect\citeauthoryear{Garn \& Alexander}{2009}]{2009MNRAS.394..105G} Garn T., Alexander P., 2009, \mnras, 394, 105 
\bibitem[\protect\citeauthoryear{Godfrey et al.}{2012}]{2012PASA...29...42G} Godfrey L.~E.~H., et al., 2012, \pasa, 29, 42 
\bibitem[\protect\citeauthoryear{Goldschmidt et al.}{1999}]{1999ApJ...511..612G} Goldschmidt P., Kukula M.~J., Miller L., Dunlop J.~S., 1999, \apj, 511, 612 
\bibitem[\protect\citeauthoryear{Greisen}{2003}]{2003ASSL..285..109G} Greisen E.~W., 2003, in Heck
A., ed., Astrophysics and Space Science Library, Vol. 285, Information Handling in Astronomy -- Historical Vistas. Kluwer, Dordrecht, p. 109
\bibitem[\protect\citeauthoryear{G{\"u}rkan et al.}{2019}]{2019AnA...622A..11G} G{\"u}rkan G., et al., 2019, \aap, 622, A11 
\bibitem[\protect\citeauthoryear{Helfand, White \& Becker}{2015}]{2015ApJ...801...26H} Helfand D.~J., White R.~L., Becker R.~H., 2015, \apj, 801, 26 
\bibitem[\protect\citeauthoryear{Hodge et al.}{2008}]{2008AJ....136.1097H} Hodge J.~A., Becker R.~H., White R.~L., de Vries W.~H., 2008, \aj 136, 1097 
\bibitem[\protect\citeauthoryear{Hooper et al.}{1996}]{1996ApJ...473..746H} Hooper E.~J., Impey C.~D., Foltz C.~B., Hewett P.~C., 1996, \apj, 473, 746 
\bibitem[\protect\citeauthoryear{Hopkins et al.}{2003}]{2003ApJ...599..971H} Hopkins A.~M., et al., 2003, \apj, 599, 971 
\bibitem[\protect\citeauthoryear{Howell et al.}{2010}]{2010ApJ...715..572H} Howell J.~H., et al., 2010, ApJ, 715, 572 
\bibitem[\protect\citeauthoryear{Hwang et al.}{2018}]{2018MNRAS.477..830H} Hwang H.-C., Zakamska N.~L., Alexandroff R.~M., Hamann F., Greene J.~E., Perrotta S., Richards G.~T., 2018, \mnras, 477, 830 
\bibitem[\protect\citeauthoryear{Hyv{\"o}nen et al.}{2007}]{2007AnA...462..525H} Hyv{\"o}nen T., Kotilainen J.~K., {\"O}rndahl E., Falomo R., Uslenghi M., 2007, \aap, 462, 525 
\bibitem[\protect\citeauthoryear{Ibar et al.}{2008}]{2008MNRAS.386..953I} Ibar E., et al., 2008, \mnras, 386, 953 
\bibitem[\protect\citeauthoryear{Jahnke, Kuhlbrodt \& Wisotzki}{2004}]{2004MNRAS.352..399J} Jahnke K., Kuhlbrodt B., Wisotzki L., 2004, \mnras, 352, 399 
\bibitem[\protect\citeauthoryear{Jiang et al.}{2010}]{2010ApJ...711..125J} Jiang Y.-F., Ciotti L., Ostriker J.~P., Spitkovsky A., 2010, \apj, 711, 125 
\bibitem[\protect\citeauthoryear{Karim et al.}{2011}]{2011ApJ...730...61K} Karim A., et al., 2011, \apj, 730, 61 
\bibitem[\protect\citeauthoryear{Kellermann et al.}{1989}]{1989AJ.....98.1195K} Kellermann K.~I., Sramek R., Schmidt M., Shaffer D.~B., Green R., 1989, \aj 98, 1195 
\bibitem[\protect\citeauthoryear{Kellermann et al.}{2016}]{2016ApJ...831..168K} Kellermann K.~I., Condon J.~J., Kimball A.~E., Perley R.~A., Ivezi{\'c} {\v Z}., 2016, \apj, 831, 168 
\bibitem[\protect\citeauthoryear{Kewley et al.}{2000}]{2000ApJ...530..704K} Kewley L.~J., Heisler C.~A., Dopita M.~A., Sutherland R., Norris R.~P., Reynolds J., Lumsden S., 2000, \apj, 530, 704 
\bibitem[\protect\citeauthoryear{Laor \& Behar}{2008}]{2008MNRAS.390..847L} Laor A., Behar E., 2008, MNRAS, 390, 847
\bibitem[\protect\citeauthoryear{Laor, Baldi \& Behar}{2019}]{2019MNRAS.482.5513L} Laor A., Baldi R.~D., Behar E., 2019, MNRAS, 482, 5513
\bibitem[\protect\citeauthoryear{Lofthouse et al.}{2018}]{2018MNRAS.479..807L} Lofthouse E.~K., Kaviraj S., Smith D.~J.~B., Hardcastle M.~J., 2018, \mnras, 479, 807 
\bibitem[\protect\citeauthoryear{Mahabal et al.}{2005}]{2005ApJ...634L...9M} Mahabal A., Stern D., Bogosavljevi{\'c} M., Djorgovski S.~G., Thompson D., 2005, \apj, 634, L9 
\bibitem[\protect\citeauthoryear{Mahajan et al.}{2019}]{2019MNRAS.482..560M} Mahajan S., Ashby M.~L.~N., Willner S.~P., Barmby P., Fazio G.~G., Maragkoudakis A., Raychaudhury S., Zezas A., 2019, \mnras, 482, 560 
\bibitem[\protect\citeauthoryear{Mahony et al.}{2012}]{2012ApJ...754...12M} Mahony E.~K., Sadler E.~M., Croom S.~M., Ekers R.~D., Feain I.~J., Murphy T., 2012, \apj, 754, 12 
\bibitem[\protect\citeauthoryear{Mao et al.}{2017}]{2017ApJ...842...87M} Mao P., Urry C.~M., Marchesini E., Landoni M., Massaro F., Ajello M., 2017, \apj, 842, 87 
\bibitem[\protect\citeauthoryear{Mart{\'{\i}}nez-Paredes et al.}{2019}]{2019ApJ...871..190M} Mart{\'{\i}}nez-Paredes M., Aretxaga I., Gonz{\'a}lez-Mart{\'{\i}}n O., Alonso-Herrero A., Levenson N.~A., Ramos Almeida C., L{\'o}pez-Rodr{\'{\i}}guez E., 2019, \apj, 871, 190 
\bibitem[\protect\citeauthoryear{Matsuoka et al.}{2016}]{2016ApJ...828...26M} Matsuoka Y., et al., 2016, \apj, 828, 26 
\bibitem[\protect\citeauthoryear{Middelberg et al.}{2011}]{2011AnA...526A..74M} Middelberg E., et al., 2011, \aap, 526, A74 
\bibitem[\protect\citeauthoryear{Miller et al.}{2008}]{2008ApJS..179..114M} Miller N.~A., et al., 2008, \apjs, 179, 114
\bibitem[\protect\citeauthoryear{Miller et al.}{2013}]{2013ApJS..205...13M} Miller N.~A., et al., 2013, \apjs, 205, 13
\bibitem[\protect\citeauthoryear{Murphy \& VLASS Survey Science Group}{2015}]{2015fers.confE...6M} Murphy E., VLASS Survey Science Group, 2015, The Many Facets of Extragalactic Radio Surveys: Towards New Scientific Challenges, Proceedings of Science, PoS(EXTRA-RADSUR2015)006
\bibitem[\protect\citeauthoryear{P{\^a}ris et al.}{2012}]{2012AnA...548A..66P} P{\^a}ris I., et al., 2012, \aap, 548, A66 
\bibitem[\protect\citeauthoryear{P{\^a}ris et al.}{2014}]{2014AnA...563A..54P} P{\^a}ris I., et al., 2014, \aap, 563, A54 
\bibitem[\protect\citeauthoryear{P{\^a}ris et al.}{2017}]{2017AnA...597A..79P} P{\^a}ris I., et al., 2017, \aap, 597, A79 
\bibitem[\protect\citeauthoryear{Paragi et al.}{2015}]{2015aska.confE.143P} Paragi Z., et al., 2015, in Advancing Astrophysics with the Square Kilometre Array, Proceedings of Science, PoS(AASKA14)143 
\bibitem[\protect\citeauthoryear{Penney et al.}{2019}]{2019MNRAS.483..514P} Penney J.~I., et al., 2019, \mnras, 483, 514 
\bibitem[\protect\citeauthoryear{Perger et al.}{2017}]{2017FrASS...4....9P} Perger K., Frey S., Gab{\'a}nyi K.~{\'E}., T{\'o}th L.~V., 2017, Front. Astron. Space Sci., 4, 9 
\bibitem[\protect\citeauthoryear{Pierce, Ballantyne \& Ivison}{2011}]{2011ApJ...742...45P} Pierce C.~M., Ballantyne D.~R., Ivison R.~J., 2011, \apj, 742, 45 
\bibitem[\protect\citeauthoryear{Pracy et al.}{2016}]{2016MNRAS.460....2P} Pracy M.~B., et al., 2016, \mnras, 460, 2
\bibitem[\protect\citeauthoryear{Rees et al.}{2016}]{2016MNRAS.455.2731R} Rees G.~A., et al., 2016, \mnras, 455, 2731
\bibitem[\protect\citeauthoryear{Reines \& Deller}{2012}]{2012ApJ...750L..24R} Reines A.~E., Deller A.~T., 2012, \apj, 750, L24 
\bibitem[\protect\citeauthoryear{Rosario et al.}{2013}]{2013ApJ...778...94R} Rosario D.~J., Burtscher L., Davies R., Genzel R., Lutz D., Tacconi L.~J., 2013, \apj, 778, 94 
\bibitem[\protect\citeauthoryear{Roy et al.}{2018}]{2018ApJ...869..117R} Roy N., et al., 2018, \apj, 869, 117 
\bibitem[\protect\citeauthoryear{Sadler et al.}{2008}]{2008ASPC..399..413S} Sadler E.~M., Johnston H.~M., Cannon R.~D., Mauch T., 2008, \aspc, 399, 413 
\bibitem[\protect\citeauthoryear{Schmitt et al.}{2006}]{2006ApJ...643..173S} Schmitt H.~R., Calzetti D., Armus L., Giavalisco M., Heckman T.~M., Kennicutt R.~C., Jr., Leitherer C., Meurer G.~R., 2006, \apj, 643, 173 
\bibitem[\protect\citeauthoryear{Shao et al.}{2019}]{2019ApJ...876...99S} Shao Y., et al., 2019, \apj, 876, 99
\bibitem[\protect\citeauthoryear{Shi et al.}{2007}]{2007ApJ...669..841S} Shi Y., et al., 2007, \apj, 669, 841 
\bibitem[\protect\citeauthoryear{Simpson et al.}{2013}]{2013MNRAS.433.2647S} Simpson C., Westoby P., Arumugam V., Ivison R., Hartley W., Almaini O., 2013, \mnras, 433, 2647 
\bibitem[\protect\citeauthoryear{Smol{\v{c}}i{\'c} et al.}{2017a}]{2017A&A...602A...1S} Smol{\v{c}}i{\'c} V., et al., 2017, \aap, 602, A1
\bibitem[\protect\citeauthoryear{Smol{\v{c}}i{\'c} et al.}{2017b}]{2017A&A...602A...2S} Smol{\v{c}}i{\'c} V., et al., 2017, \aap, 602, A2
\bibitem[\protect\citeauthoryear{Smol{\v{c}}i{\'c} et al.}{2017c}]{2017A&A...602A...6S} Smol{\v{c}}i{\'c} V., et al., 2017, \aap, 602, A6 
\bibitem[\protect\citeauthoryear{Strittmatter et al.}{1980}]{1980AnA....88L..12S} Strittmatter P.~A., Hill P., Pauliny-Toth I.~I.~K., Steppe H., Witzel A., 1980, \aap, 88, L12 
\bibitem[\protect\citeauthoryear{Ulvestad \& Ho}{2002}]{2002ApJ...581..925U} Ulvestad J.~S., Ho L.~C., 2002, \apj, 581, 925 
\bibitem[\protect\citeauthoryear{Venemans et al.}{2016}]{2016ApJ...816...37V} Venemans B.~P., Walter F., Zschaechner L., Decarli R., De Rosa G., Findlay J.~R., McMahon R.~G., Sutherland W.~J., 2016, \apj, 816, 37 
\bibitem[\protect\citeauthoryear{Venemans et al.}{2017}]{2017ApJ...837..146V} Venemans B.~P., et al., 2017, \apj, 837, 146 
\bibitem[\protect\citeauthoryear{Venemans et al.}{2019}]{2019ApJ...874L..30V} Venemans B., Neeleman M., Walter F., Novak M., Decarli R., Hennawi J., Rix H.-W., 2019, \apj, 874, L30
\bibitem[\protect\citeauthoryear{Wals et al.}{2005}]{2005MNRAS.360..453W} Wals M., Boyle B.~J., Croom S.~M., Miller L., Smith R., Shanks T., Outram P., 2005, \mnras, 360, 453 
\bibitem[\protect\citeauthoryear{Walter et al.}{2004}]{2004ApJ...615L..17W} Walter F., Carilli C., Bertoldi F., Menten K., Cox P., Lo K.~Y., Fan X., Strauss M.~A., 2004, \apj, 615, L17 
\bibitem[\protect\citeauthoryear{Wang et al.}{2013}]{2013ApJ...773...44W} Wang R., et al., 2013, \apj, 773, 44 
\bibitem[\protect\citeauthoryear{Wang et al.}{2016}]{2016ApJ...830...53W} Wang R., et al., 2016, \apj, 830, 53 
\bibitem[\protect\citeauthoryear{Wang et al.}{2017}]{2017ApJ...835L..20W} Wang R., et al., 2017, \apj, 835, L20 
\bibitem[\protect\citeauthoryear{Warren et al.}{1987}]{1987Natur.325..131W} Warren S.~J., Hewett P.~C., Irwin M.~J., McMahon R.~G., Bridgeland M.~T., 1987, \nat, 325, 131 
\bibitem[\protect\citeauthoryear{Wells, Greisen \& Harten}{1981}]{1981AnAS...44..363W} Wells D.~C., Greisen E.~W., Harten R.~H., 1981, \aaps, 44, 363 
\bibitem[\protect\citeauthoryear{White et al.}{2007}]{2007ApJ...654...99W} White R.~L., Helfand D.~J., Becker R.~H., Glikman E., de Vries W., 2007, \apj, 654, 99 
\bibitem[\protect\citeauthoryear{Willott, Bergeron \& Omont}{2015}]{2015ApJ...801..123W} Willott C.~J., Bergeron J., Omont A., 2015, \apj, 801, 123 
\bibitem[\protect\citeauthoryear{Willott, Omont \& Bergeron}{2013}]{2013ApJ...770...13W} Willott C.~J., Omont A., Bergeron J., 2013, \apj, 770, 13 
\bibitem[\protect\citeauthoryear{Willott et al.}{1998}]{1998MNRAS.300..625W} Willott C.~J., Rawlings S., Blundell K.~M., Lacy M., 1998, \mnras, 300, 625 
\bibitem[\protect\citeauthoryear{Wright et al.}{2010}]{2010AJ....140.1868W} Wright E.~L., et al., 2010, \aj 140, 1868 
\bibitem[\protect\citeauthoryear{Wu et al.}{2005}]{2005ApJ...632L..79W} Wu H., Cao C., Hao C.-N., Liu F.-S., Wang J.-L., Xia X.-Y., Deng Z.-G., Young C.~K.-S., 2005, \apj, 632, L79 
\bibitem[\protect\citeauthoryear{York et al.}{2000}]{2000AJ....120.1579Y} York D.~G., et al., 2000,\aj 120, 1579 
\bibitem[\protect\citeauthoryear{Yun, Reddy \& Condon}{2001}]{2001ApJ...554..803Y} Yun M.~S., Reddy N.~A., Condon J.~J., 2001, \apj, 554, 803 
\bibitem[\protect\citeauthoryear{Zakamska \& Greene}{2014}]{2014MNRAS.442..784Z} Zakamska N.~L., Greene J.~E., 2014, \mnras, 442, 784 
\bibitem[\protect\citeauthoryear{Zakamska et al.}{2016}]{2016MNRAS.455.4191Z} Zakamska N.~L., et al., 2016, \mnras, 455, 4191 
\bibitem[\protect\citeauthoryear{Zamfir, Sulentic \& Marziani}{2008}]{2008MNRAS.387..856Z} Zamfir S., Sulentic J.~W., Marziani P., 2008, \mnras, 387, 856 
\bibitem[\protect\citeauthoryear{Zensus}{1997}]{1997ARAnA..35..607Z} Zensus J.~A., 1997, \araa, 35, 607 

\end{thebibliography}
\end{document}